 \documentclass[final,3p,times]{elsarticle}
\usepackage{amssymb}
\usepackage{amsmath}
\usepackage{textcomp}
\usepackage[font=footnotesize, margin=3em]{caption}
\usepackage[T1]{fontenc}
\usepackage{lmodern}  %% for a fast and nice pdf when viewed on screen
\usepackage[utf8]{inputenc}
\usepackage{color}
\usepackage{pdflscape} % Oldal elforgatásához
\usepackage{gensymb} % A degree parancshoz
\usepackage{float} % Képek rögzítéséhez
\usepackage{verbatim} % blokkok kikommenteléséhez
\usepackage{hyperref}
\usepackage{csquotes}
\usepackage{subfigure}
\hypersetup{
	colorlinks=true,
	linkcolor=black,
	citecolor=black,
}

{ % táblázat vertikális nyújtása

\def\f{\frac}

\journal{International Journal of Heat and Mass Transfer}

\begin{document}

\begin{frontmatter}



\title{Second sound and ballistic heat conduction: NaF experiments revisited}


\author{R. Kov\'acs$^{123}$ and P. V\'an$^{123}$}

\address{
$^1$Department of Theoretical Physics, Wigner Research Centre for Physics,
Institute for Particle and Nuclear Physics, Budapest, Hungary and
$^2$Department of Energy Engineering, BME, Budapest, Hungary and
$^3$Montavid Thermodynamic Research Group}

\begin{abstract}
Second sound phenomenon and ballistic heat conduction, the two wave like propagation modes of heat, are the two most prominent, experimentally observed non-Fourier effects of heat conduction. In this paper we compare three related theories by quantitatively analyzing the crucial NaF experiments of Jackson, Walker and McNelly, where these effects were observed together. We conclude that with the available information the best comparison and insight is provided by non-equilibrium thermodynamics with internal variables. However, the available data and information is not the best, and further, new experiments are necessary.

\end{abstract}


\end{frontmatter}


\section{Introduction}
Due to the technological development, manufacturing and material designing achieved the level where the classical laws of physics and the related engineering methodologies do not hold. The challenging areas are the low temperatures and nano-scales. This is most relevant for heat conduction where the deviation from the classical Fourier law are well known since decades but the circumstances leading to non-Fourier phenomena are not yet clear. In particular the discovery of Guyer-Krumhansl-type heat conduction in heterogeneous materials \cite{Botetal16, Vanetal17} at room temperature shows, that the traditional view of the validity of these theories is too narrow, and requires further investigations. This is also important for the two prominent non-Fourier phenomena, for the second sound and for the ballistic propagation. 

The second sound denotes the wave like propagation of heat modelled by the Maxwell-Cattaneo-Vernotte (MCV) equation. Here the propagation speed of heat waves is material dependent, but less than the speed of sound wave. In case of  ballistic conduction the heat propagates exactly at the speed of mechanical sound waves. The later one can be understood and interpreted in different ways depending on the underlying theory. 

In phonon hydrodynamics and in particular in Rational Extended Thermodynamics \cite{DreStr93a, MulRug98}, ballistic conduction is understood as a non-interactive propagation of phonons. They are reflected and scattered only on the boundaries. In the framework of non-equilibrium thermodynamics with internal variables \cite{KovVan15}, the ballistic propagation is represented as a coupling between the thermal and mechanical fields and propagates with the elastic wave. In the complex viscosity model of Rogers it is a phenomenon of pure mechanical origin \cite{Rog71a}. The propagation speed is common in these three theories, it is the speed of sound. The experimental detection of second sound and ballistic modes is not easy and requires a theoretical insight, too. The theoretical predictions of second sound by Tisza and Landau have been based on their respective two fluid theories \cite{Tisza38, Lan41}. It was measured first by Peshkov \cite{Pesh44} in super-fluid He-4. The so-called window condition derived by Guyer and Krumhansl \cite{GK66} indicated the frequency range, where the dissipation of the wave propagation is minimal and significantly aided the detection of second sound in solids. However, such kind of theoretical tool for ballistic type conduction does not exist yet. Moreover, it is not clear what really influences the existence of ballistic signals \cite{KovVan16}.  

\begin{figure}[h]
\centering
\includegraphics[width=8.5cm,height=7cm]{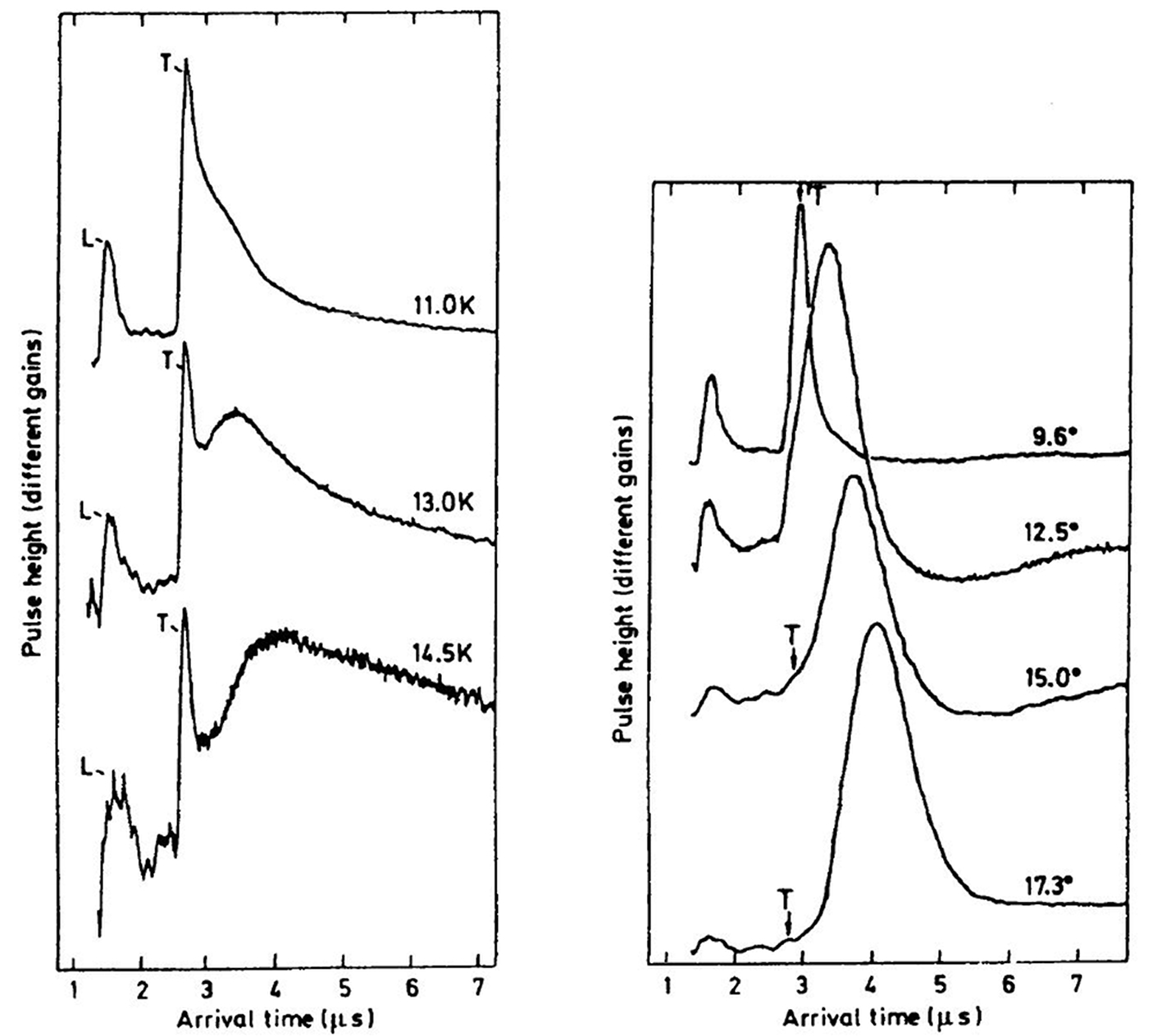}
\caption{NaF experiment results: the left one ("\#607167J") was published in \cite{JacWalMcN70} and the right one ("\#7204205W") in \cite{JacWal71}.}
\label{fig:bal1}
\end{figure}

Let us shortly review the essential information of experiments of Jackson, Walker and McNelly \cite{JacWalMcN70,JacWal71,McN74t}. According to McNelly's PhD thesis \cite{McN74t}, each NaF sample is identified by a number. Here measurements on two different NaF crystals are analyzed, these are called as "\#607167J" and "\#7204205W" \cite{McN74t}. Figure \ref{fig:bal1} shows the experimental results related to these samples.

\begin{figure}[h]
\centering
\includegraphics[width=8.5cm,height=7cm]{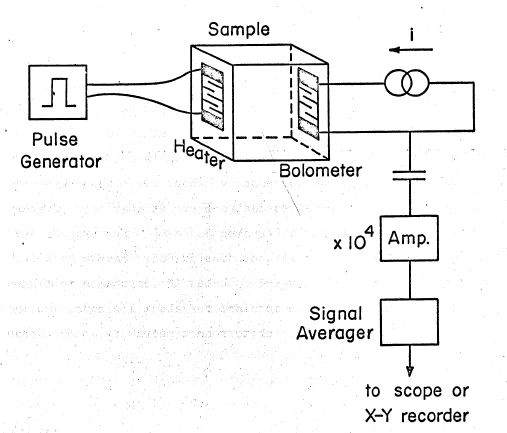}
\caption{Arrangement of NaF experiment, original figure from \cite{McN74t}.}
\label{fig:McN1}
\end{figure}

These curves on Fig. \ref{fig:bal1} correspond to a so-called heat pulse experiment. Their schematic arrangement is presented on Fig. \ref{fig:McN1} \cite{McN74t}. The rear side temperature history is measured and presented previously on Fig. \ref{fig:bal1}.

In the next section we shortly review the relevant theories. Then we collect and compare the performance of these theories considering both qualitative and quantitative aspects.

\section{Models of ballistic heat conduction}

Here we mention three different approaches to describe ballistic type propagation:
\begin{itemize}
\item kinetic theory based phonon hydrodynamics of Dreyer and Struchtrup \cite{DreStr93a, MulRug98},
\item complex viscosity based hybrid phonon gas model of Y. Ma \cite{Ma13a},
\item non-equilibrium thermodynamics with internal variables and Nyíri-multipliers \cite{KovVan15, KovVan16}.
\end{itemize}
These models provided results regarding the reproduction of NaF experiments. One should mention also the work of Frischmuth and Cimmelli based on thermoelastic approach \cite{FriCim95}, Cimmelli et al. \cite{Cimm09diff, Cimmelli09nl, JouCimm16} and Bargmann and Steinmann \cite{BarSte05a, BarSte08a} in spite of the poor reproduction of the measured results.

\subsection{Kinetic theory and phonon hydrodynamics}

It is a particle based approach with phonons. The interactions between phonons can be classified as \cite{DreStr93a, MulRug98}:
\begin{itemize}
\item Normal (N) processes: the total momentum of phonons is conserved during the interaction,
\item Resistive (R) processes: the opposite of N processes, total momentum is not conserved.
\end{itemize}
Common in both cases is that the energy is conserved during the interactions. A third type, called as Umklapp process also exists where neither the energy nor the momentum is conserved  \cite{MulRug98}. The thermal conductivity $\lambda$ is expected to be in connection with R processes and their frequencies $\frac{1}{\tau_R}$, i.e.
\begin{equation}
\lambda=\frac{c^2}{3} c_v \tau_R ,
\end{equation}
where $c$ is the Debye speed of phonons, $c_v$ is the isochoric specific heat, and $\tau_R$ is the characteristic time of the R processes.
In this approach,  Fourier's law can be applied only when the R processes are significantly dominant which results the diffusive kind of heat propagation. However, if one decreases the temperature then N processes become dominant and the wave nature of heat conduction reveals itself. In case of ballistic propagation there is no interaction between phonons, these particles just go through the sample without colliding. In order to include these propagation modes into phonon hydrodynamics, one needs to distinguish between the phase density for the R and N processes. The resistive processes tend to the function $f_R$ and the normal processes tend to the distribution function $f_N$:
\begin{eqnarray}
f_R&=&\frac{y}{exp \left(\frac{hck}{k_BT} \right) -1}, \\
f_N&=&\frac{y}{exp \left \{ \frac{hck}{k_BT} \left(1- \frac{3}{4} \frac{cp_in_i}{aT^4} \right) \right \} -1},
\end{eqnarray}
where $h$ is the Planck constant, $k_B$ is the Boltzmann constant, $k$ denotes the wavenumber, $n_i$ is the unit vector in the direction of $k$, $p_i$ is the momentum, $y = 3/(8 \pi^3)$ is  a constant, $a$ is also a constant, see below. The temperature $T$ is defined as the energy density by Debye law for phonons \cite{DreStr93a}:
\begin{equation}
e=aT^4; \ \ a=\frac{4\pi^5}{5}\frac{k_B^4}{h^3c^3}.
\end{equation}
The phase density $f$ of one phonon evolves according to the Boltzmann equation,
\begin{equation}
\partial_t f + c n_i \partial_i f = \hat{S},
\end{equation}
where $\hat{S}$ is the collision integral. In the Callaway model the previous equilibrium distributions $f_R$ and $f_N$ are considered and combined in relaxation terms:
\begin{equation}
\hat{S} = - \f{1}{\tau_R} ( f - f_R) - \f{1}{\tau_N} ( f - f_N) .
\end{equation}
This assumption implies that two different equilibrium distributions exist in this system.
Instead of solving the Boltzmann equation, one can approximate the solution by momentum series expansion. It leads to a system of momentum equations and introduce new quantities this way:
\begin{equation}
u_{\langle i_1i_2...i_N \rangle}=\int kn_{\langle i_1...i_n \rangle}f dk.
\end{equation}
Here $_{\langle \ \rangle}$ denotes the traceless symmetric part of a tensor. The first momentum is the energy density, the second one is the momentum density, the third one is the energy flux and the fourth one is the deviatoric part of the pressure tensor \cite{MulRug98}, i.e.
\begin{equation}
e=hcu; \ \ p_i=hu_i; \ \ Q_i=hc^2u_i; \ \ N_{\langle ij \rangle}=hcu_{\langle ij \rangle} .
\end{equation}
Let us note that this method leads to a system with infinite number of equations and leads to the closure problem. Here for heat conduction, one applies the simplest one, neglecting the highest order flux, truncating the series, i.e. the new highest order (``$N+1$''$^{\rm{th}}$) quantity is considered as zero. One obtains the following system of partial differential equations in 1+1 dimensions:
\begin{equation}
\frac{\partial u_{\langle n \rangle}}{\partial t} + \frac{n^2}{4n^2-1}c\frac{\partial u_{\langle n-1 \rangle}}{\partial x}+c\frac{\partial u_{\langle n+1 \rangle}}{\partial x}=\left \{ \begin{array}{ll}
\displaystyle
0 \ & \ n=0 \\
-\frac{1}{\tau_R}u_{\langle 1 \rangle} \ & \ n=1 \\
- \left( \frac{1}{\tau_R}+\frac{1}{\tau_N} \right )u_{\langle n \rangle} \ & \ 2\leq n\leq N
\end{array} \right. \label{phhysys}
\end{equation}
One has to apply $N \cong 30$ equations to obtain a good approximation of the ballistic propagation speed of phonons. Naturally, it is  difficult to solve (\ref{phhysys}) for practical problems, but $N=3$ equations give an acceptable approximation, while one accepts that the predicted value of the ballistic propagation speed is not correct. For $N=3$ one obtains a 3 field theory:
\begin{eqnarray}
\partial_t e + c^2 \partial_x p &=& 0, \nonumber \\
\partial_t p + \frac{1}{3} \partial_x e +\partial_x N &=& -\frac{1}{\tau_R} p, \label{eq_kt_ball} \\
\partial_t N + \frac{4}{15} c^2 \partial_x p &=& - \left( \frac{1}{\tau_R}+\frac{1}{\tau_N} \right ) N. \nonumber
\end{eqnarray}

\subsection{Hybrid phonon gas model}

Ma developed this approach based on the work of Rogers \cite{Rog71a} and Landau \cite{LandauVIeng} to describe the longitudinal and transversal ballistic signals at the same time \cite{Ma13a, Ma13a1}. In this model, the internal energy $E$ is splitted into two parts:
\begin{equation}
E = E_0 + E',
\end{equation}
where $E_0$ corresponds to the equilibrium part and $E'$ is the perturbation. It is supposed to consist of the longitudinal and transversal parts as
\begin{equation}
E'=E'_l + 2E'_t.
\end{equation}
Let us consider the classical equation of motion of a viscous fluid:
\begin{equation}
\rho \partial_t \mathbf{v} + \rho( \mathbf{v} \cdot grad) \mathbf{v} = - grad P + \eta \nabla^2 \mathbf{v} + (\xi + \frac{1}{3} \eta) grad \ div \mathbf{v},
\label{cns}
\end{equation}
where $\mathbf{v}$, $P$, $\rho$, $\xi$ and $\eta$ are the velocity, pressure, mass density, shear and bulk viscosities, respectively.
As Rogers stated in \cite{Rog71a}, for phonon gas the equation (\ref{cns}) is valid for heat flux $\mathbf{q} = E \cdot \mathbf{v}$, too. Moreover, when relaxation time $\tau_N$ related to the normal processes is increasing the shear viscosity tends to zero and only the bulk term plays a role in the damping mechanism \cite{Rog71a}. Analogously with the hydrodynamic case \cite{LandauVIeng}, the bulk viscosity $\xi$ is rewritten as
\begin{equation}
\xi = \frac{\tau E (1 - c_2^2 / c_1 ^2)}{1 - i \omega \tau},
\end{equation}
where $\tau^{-1} = \tau_R^{-1} + \tau_N^{-1}$ and $c_1, c_2$ are the characteristic first and second sound velocities. Then it leads to the system related to the transversal propagation mode (denoted by subscript t) reduced to one dimension:
\begin{eqnarray}
\partial_t E'_t + \partial_x q_t &=& 0, \nonumber \\
\partial_t q_t + \frac{1}{3} \partial_x E'_t &=& - \frac{1}{\tau_R} q_t + \frac{2 \tau}{3 ( 1 - i \omega \tau)} \partial^2_x q_t.
\end{eqnarray}
The longitudinal part has the same form but scaled with the ratio of the propagation speeds $c_t / c_l$.

\subsection{Non-equilibrium thermodynamics}

In non-equilibrium thermodynamics internal variables and current multipliers provide the necessary extension beyond local equilibrium \cite{Ver97b,Nyi91a1,Van01a2,BerVan17b}. In \cite{KovVan15}, the theoretical model is presented and the ballistic-conductive model (BC) is derived. This model consists of the same terms as the one based on phonon hydrodynamics \cite{DreStr93a}, but the coefficients are different. The ballistic-conductive (BC) model in one spatial dimension is
\begin{eqnarray}
\rho c \partial_t T + \partial_x q & = & - a (T - T_0) \nonumber, \\
\tau_q \partial_t q +  q + \lambda \partial_x T + \kappa_{21} \partial_x Q &=& 0  \nonumber , \\
\tau_Q \partial_t Q +  Q - \kappa_{12} \partial_x q&=& 0 , \label{bc_eq3}
\end{eqnarray}
where $q$ is the heat flux, $Q$ is the current density of heat flux, $T$ stands for the temperature, $\rho$ and $ c$ are the mass density and specific heat, $\tau_q$ and $\tau_Q$ are the relaxation times corresponding to the respective fields, $\kappa_{21}$ and $\kappa_{12}$ form the antisymmetric part of the corresponding Onsager conductivity matrix, i.e. $\kappa_{21} = - \kappa_{12}$ and called dissipation parameter \cite{KovVan15}. The coefficient $\lambda$ is the thermal conductivity, $a$ is the volumetric heat transfer coefficient introduced in \cite{KovVan16} to model the internal cooling of the wave propagation channel. According to the experimental setup shown on Fig. \ref{fig:McN1} both the nonuniform heating of the front end and the cooling on the sides may be relevant modeling conditions. For heat pulse experiments the following dimensionless quantities are introduced \cite{KovVan15}:
\begin{eqnarray}
\hat{t} =\frac{\alpha t}{L^2} \quad &\text{with}& \quad
\alpha=\frac{\lambda}{\rho c};  \quad
\hat{x}=\frac{x}{L};\nonumber \\
\hat{T}=\frac{T-T_{0}}{T_{\text{end}}-T_{0}} \quad &\text{with}&\quad
T_{\text{end}}=T_{0}+\frac{\bar{q}_0 t_p}{\rho c L};  \nonumber \\
\hat{q}=\frac{q}{\bar{q}_0} \quad &\text{with}&\quad
\bar{q}_0=\frac{1}{t_p}  \int_{0}^{t_p} q_{0}(t)dt; \nonumber \\
%\hat{Q} = \sqrt{- \f{\kappa_{12}}{\kappa_{21}}} \bar{q}_0 Q, \quad &\text{and}& \quad \hat h = h \f{t_p}{\rho c L}, \nonumber \\
\hat{Q} = \sqrt{- \f{\kappa_{12}}{\kappa_{21}}} \bar{q}_0 Q, \quad &\text{and}& \quad \hat h = h \f{t_p}{\rho c L}, \quad \hat a =\frac{t_p}{\rho c} a,
\label{ndvar}\end{eqnarray}
where $\alpha$ is the thermal diffusivity, $t_p$ is the length of the heat pulse, $T_{end}$ is the maximum or equilibrium temperature value in the adiabatic case, $\bar q$ is the mean value of the heat pulse. Furthermore, the following dimensionless parameters are also introduced, correspondingly
\begin{equation}
\hat{\tau}_\Delta =\frac{\alpha t_p}{L^2}; \quad
\hat{\tau}_q 	  = \frac{\alpha \tau_{q}}{L^2}; \quad
\hat{\tau}_Q 	  = \frac{\alpha  \tau_{Q}}{L^2}; \quad
\hat{\kappa} 	  = \f{\sqrt{-\kappa_{12} \kappa_{21}}}{L},
\end{equation}
where $\hat{\tau}_\Delta$ is the dimensionless pulse length, $\hat{\tau}_q$ and $\hat{\tau}_Q$ are the dimensionless relaxation times and $\hat{\kappa}$ is the dimensionless square root of the dissipation parameter.
One obtains the dimensionless form of ballistic-conductive (BC) model \cite{KovVan15}:
\begin{eqnarray}
\hat{\tau}_{\Delta}\partial_{\hat t} \hat T +
	\partial_{\hat x} \hat q &=& - \hat a \hat T ,\nonumber \\
\hat{\tau}_q \partial_{\hat t} \hat q + \hat q +
    \hat{\tau}_{\Delta}\partial_{\hat x}\hat T +
    \hat{\kappa}\partial_{\hat x}\hat Q &=& 0
    , \nonumber \\
\hat{\tau}_Q \partial_{\hat t}\hat Q +\hat Q +
	\hat \kappa \partial_{\hat x}\hat q &=& 0 . \label{nd_balcond}
\end{eqnarray}
It is solved numerically in a way described in \cite{KovVan15}.
Its first test is presented in \cite{KovVan16}, where only one measurement is simulated, namely the one on "\#607167J" corresponding to 13 K. The main conclusions of \cite{KovVan16} are the following:
\begin{itemize}
\item The samples are ambiguously identified in \cite{JacWalMcN70, JacWal71} and it leads to misunderstandings regarding the material parameters.
\item The role of boundary conditions must be emphasized. Concerning the heat pulse, its length is not described anywhere, only an interval is mentioned by Jackson et al., i.e. somewhere between $0.1 \mu s$ and $1 \mu s$. We applied the values given by Y. Ma \cite{Ma13a1, Ma13a}.
\item The presence of cooling should be considered due to a point like excitation on the front end. It is accounted as a heat transfer term in the balance equation of internal energy. Equivalently, Dreyer and Structhtup \cite{DreStr93a} solves their phonon hydrodynamical model on semi-infinite region instead of finite domain to obtain such decreasing characteristic.
\end{itemize}
This experience regarding the modeling of NaF experiments is necessary to reproduce the other measurements, too.

%---------------------------------------------------------------------------------------------------
\subsection{Boundary and initial conditions}
As a front side boundary condition a smooth heat pulse is defined with dimensionless quantities:
\begin{center}
	$q(\hat x=0,\hat t)= \left\{ \begin{array}{cc}
	\left(1-cos\left(2 \pi \cdot \frac{\hat t}{\hat t_p}\right)\right) &
	\textrm{if } 0< \hat t \leq \hat t_p,\\
	0 & \textrm{if } \hat t> \hat t_p,
            \end{array} \right.  $
\end{center}
its length is different in the two series of experiments \cite{Ma13a1, Ma13a}. On the rear side we used only the adiabatic boundary condition ($q(\hat x=1,\hat t)=0$) because the bolometer measures the signal directly at this point. Regarding the initial conditions, all fields are homogeneously zero at the initial time instant. It means homogeneous temperature distribution which is equal to the reference temperature of each crystal.

\section{Material parameters}

This is one of the most crucial part to simulate ballistic heat conduction. McNelly's PhD thesis \cite{McN74t} is used to find and interpolate the proper value of thermal conductivity. The data of thermal conductivity from \cite{JacWalMcN70, JacWal71} are not used. Moreover, the specific heat according the paper of Hardy and Jaswal \cite{HarJas71a} is calculated for each reference temperature value. Due to the lack of temperature dependence of mass density we used its value corresponding to @15K \cite{BarSte05a, Gmelin93}.

Tables \ref{naf:matpar1} and \ref{naf:matpar2} sum up the classical material parameters used in our calculations.

\begin{table}[H]
\centering
\begin{tabular}{c|c|c|c}
       &Thermal conductivity $\left [\frac{W}{mK} \right ]$ & Specific heat $\left [\frac{J}{kgK} \right ]$  &  Mass density  $\left [\frac{kg}{m^3} \right ]$ \\ \hline
@11 K & 8573 & 1.118 & 2866  \\ \hline
@13 K &10200 & 1.8 & 2866\\ \hline
@14.5 K &10950& 2.543 & 2866\\
    \end{tabular}\\
\caption{Classical material parameters for crystal \#607167J}
\label{naf:matpar1}
\end{table}

\begin{table}[H]
\centering
\begin{tabular}{c|c|c|c}
       &Thermal conductivity $\left [\frac{W}{mK} \right ]$ & Specific heat $\left [\frac{J}{kgK} \right ]$  &  Mass density  $\left [\frac{kg}{m^3} \right ]$ \\ \hline
@9.6 K & 8500& 0.7123 & 2866  \\ \hline
@12.5 K &17300 & 1.62 & 2866\\ \hline
@15 K &21750 & 2.735 & 2866\\ \hline
@17.3 K &22880& 4.45 & 2866\\
 \end{tabular}\\
\caption{Classical material parameters for crystal \#7204205W}
\label{naf:matpar2}
\end{table}

The missing relaxation times are determined from simulations, i.e. fitted for the corresponding measurement along with the volumetric heat transfer coefficient. When comparing the ballistic-conductive model to the phonon hydrodynamical one, the difference is that the parameter $\kappa$ is a free one but in the kinetic theory it is not. A crucial property of phonon hydrodynamical model is that, one has to apply at least 30 momentum equation to approximate the ballistic propagation speed. In the BC model, one can adjust $\kappa$ to obtain the proper propagation speed based on the characteristic speed $\hat c$ \cite{KovVan15, KovVan16}:
\begin{equation}
\hat{c} = \sqrt{\frac{\hat{\kappa}^2 + \hat{\tau_Q}}{\hat{\tau_q} \hat{\tau_Q}}}.
\label{hp1}\end{equation}

\section{Results}
One should be aware of the fact that there is no vertical scale on the original measurement results (Fig. \ref{fig:bal1}). In order to overcome this shortcoming, two constraints are introduced. The first one accounts the relative amplitudes of each propagation mode, as it is introduced in \cite{KovVan16}. The second one concerns the relaxation times. It is assumed that the ratio of $\tau_q$ and $\tau_Q$ can not change between two measurements too much, i.e. order of magnitudes. These constraints beside the arrival of each signal seem to be enough for complete reproduction.

\begin{figure}[H]
     \begin{center}
        \subfigure[Related curve: @11K, sample \#607167J]{%
            \label{fig:naf_sim_1}
            \includegraphics[width=7.5cm,height=4cm]{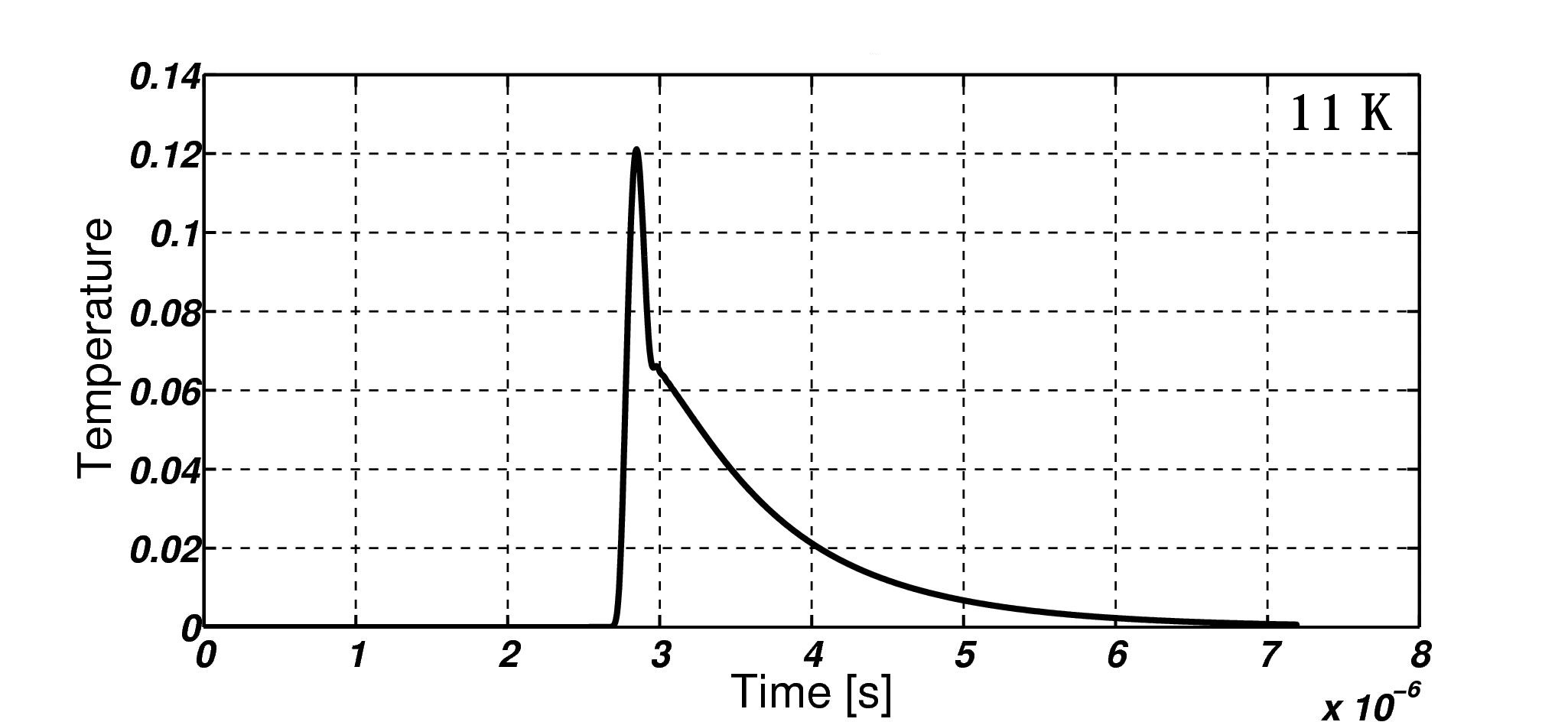}
        }%
        \subfigure[Related curve: @13K, sample \#607167J]{%
           \label{fig:naf_sim_2}
           \includegraphics[width=7.5cm,height=4cm]{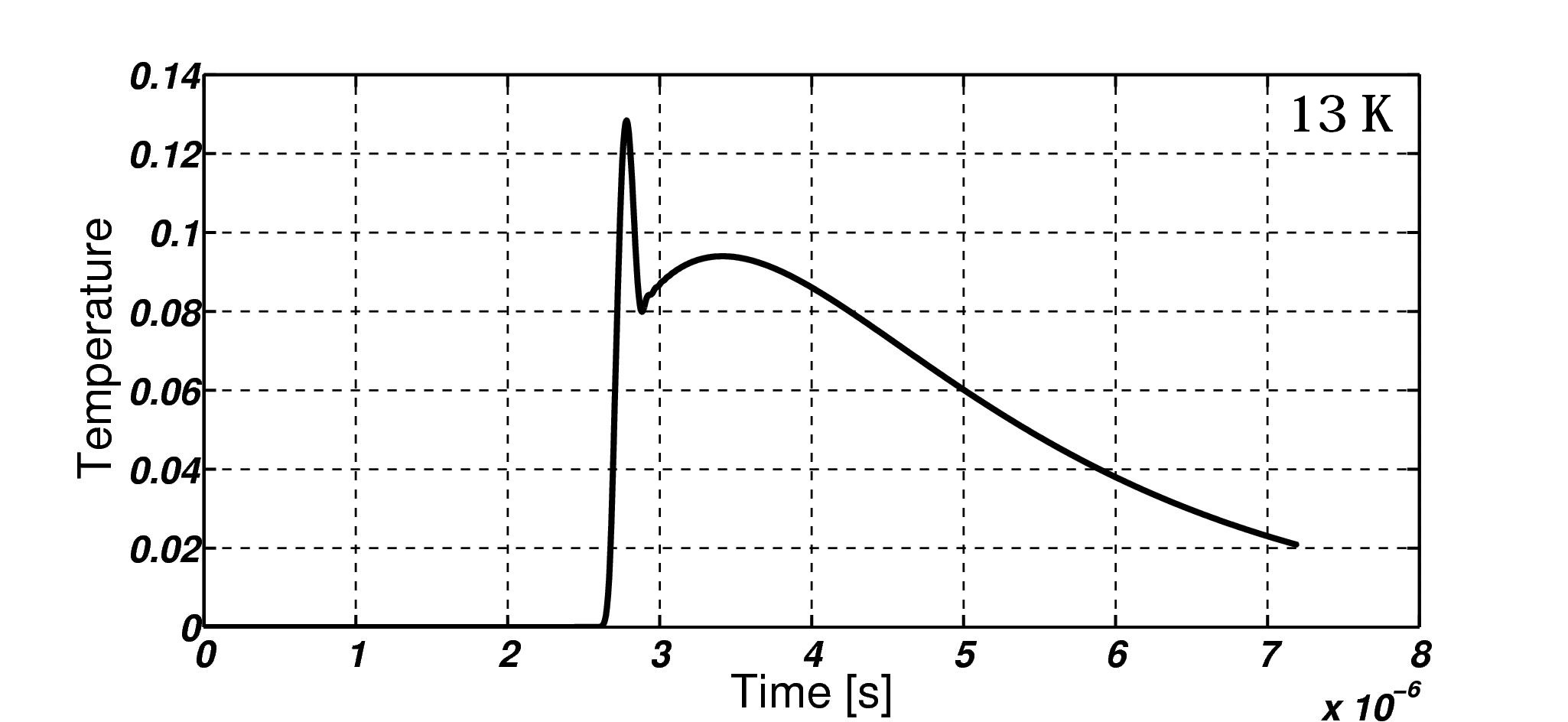}
        } %
        \subfigure[Related curve: @14.5K, sample \#607167J]{%
            \label{fig:naf_sim_3}
            \includegraphics[width=7.5cm,height=4cm]{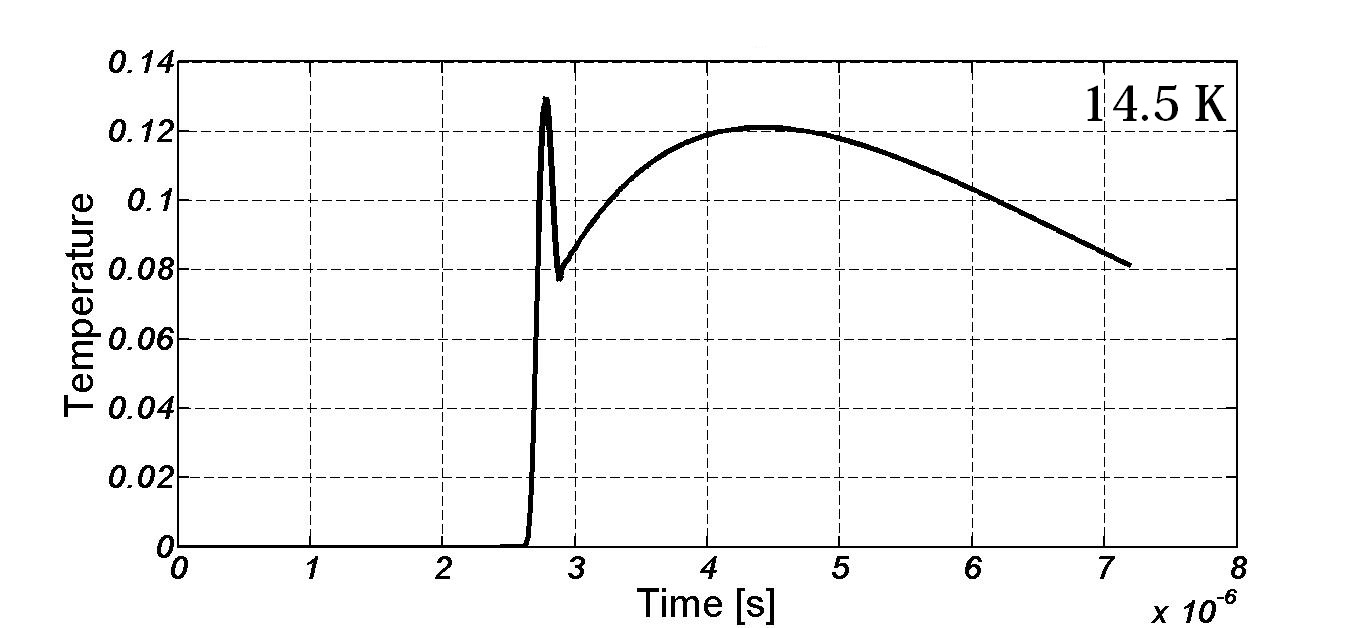}
        }%
         \subfigure[Related curve: @9.5K, sample \#7204205W]{%
            \label{fig:naf_sim_4}
            \includegraphics[width=7.5cm,height=4cm]{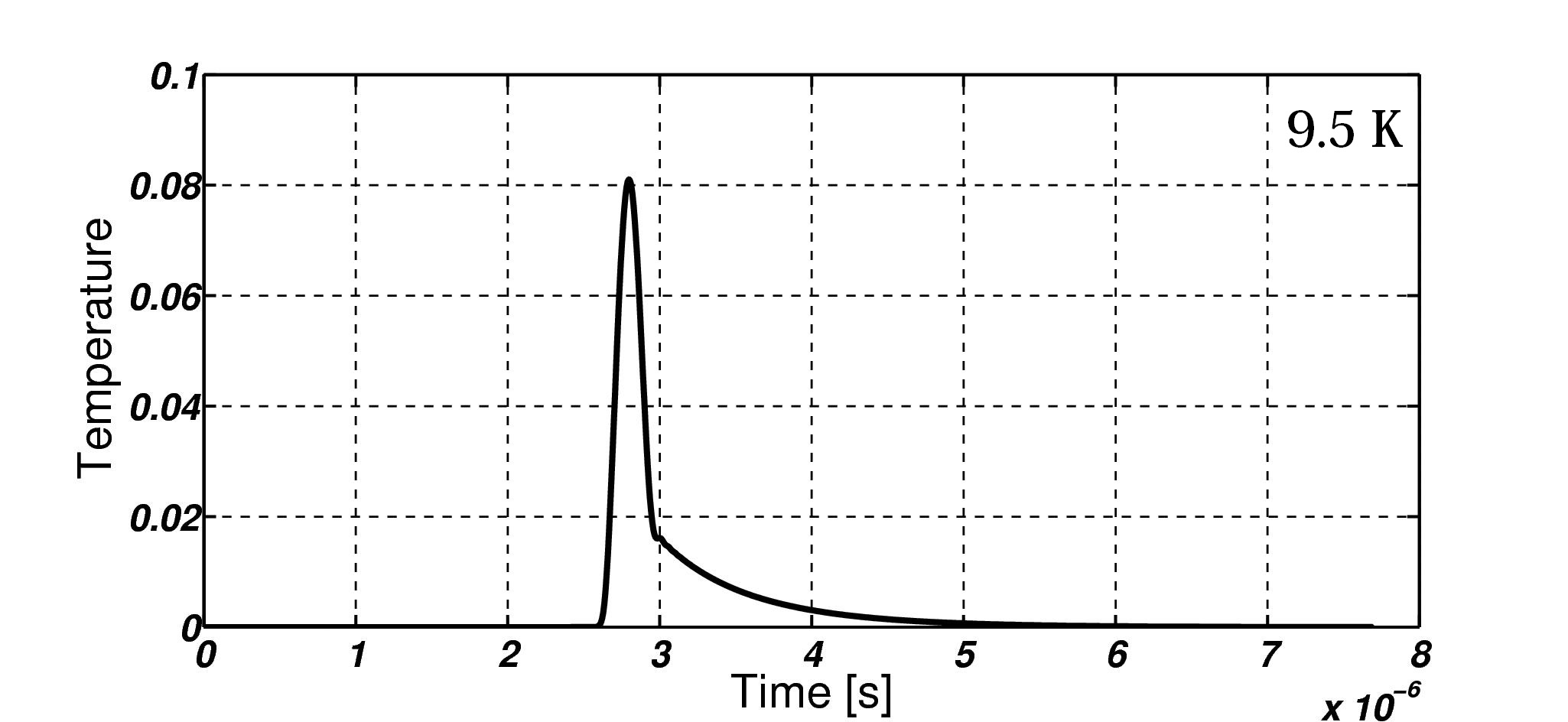}
        }% 
        \\ \subfigure[Related curve: @12.5K, sample \#7204205W]{%
            \label{fig:naf_sim_5}
            \includegraphics[width=7.5cm,height=4cm]{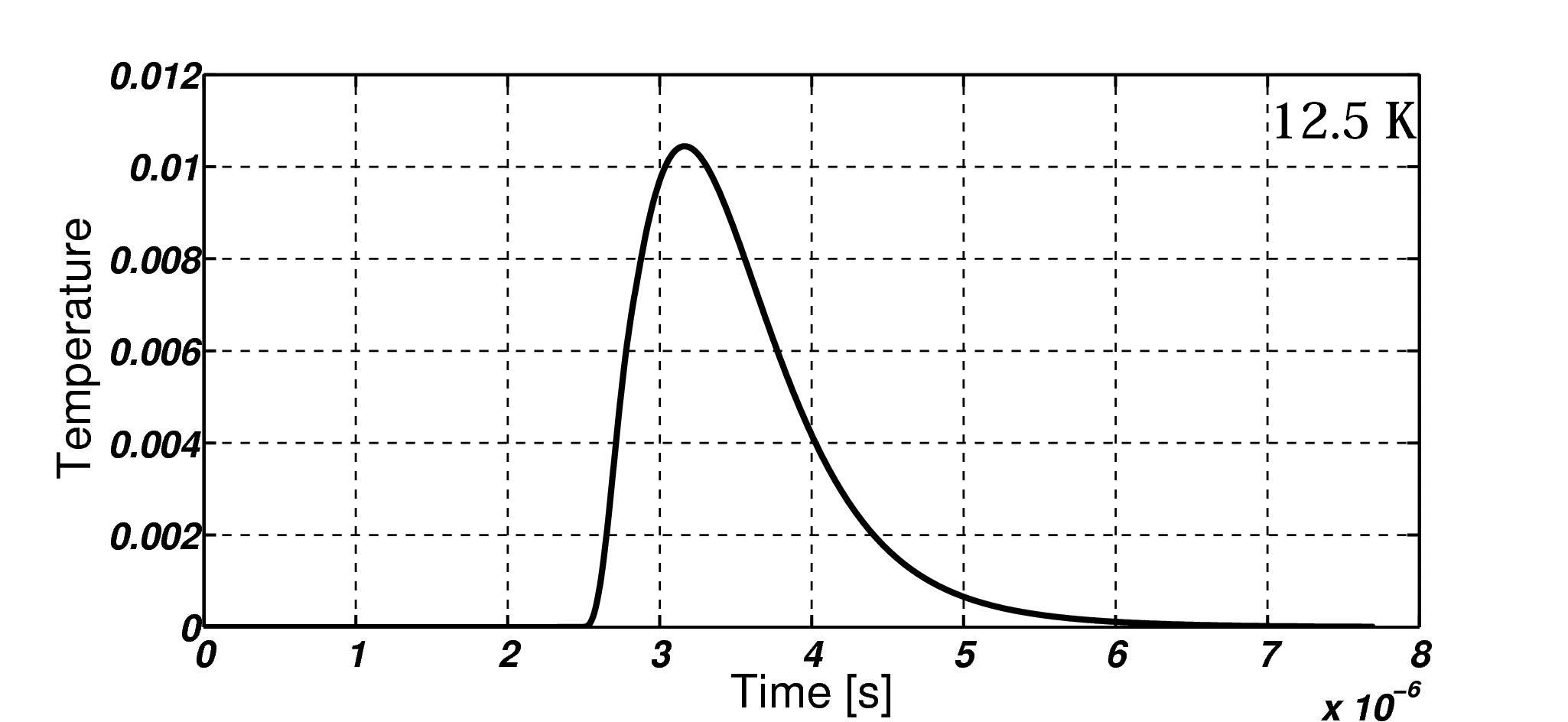}
        }% 
         \subfigure[Related curve: @15K, sample \#7204205W]{%
           \label{fig:naf_sim_6}
            \includegraphics[width=7.5cm,height=4cm]{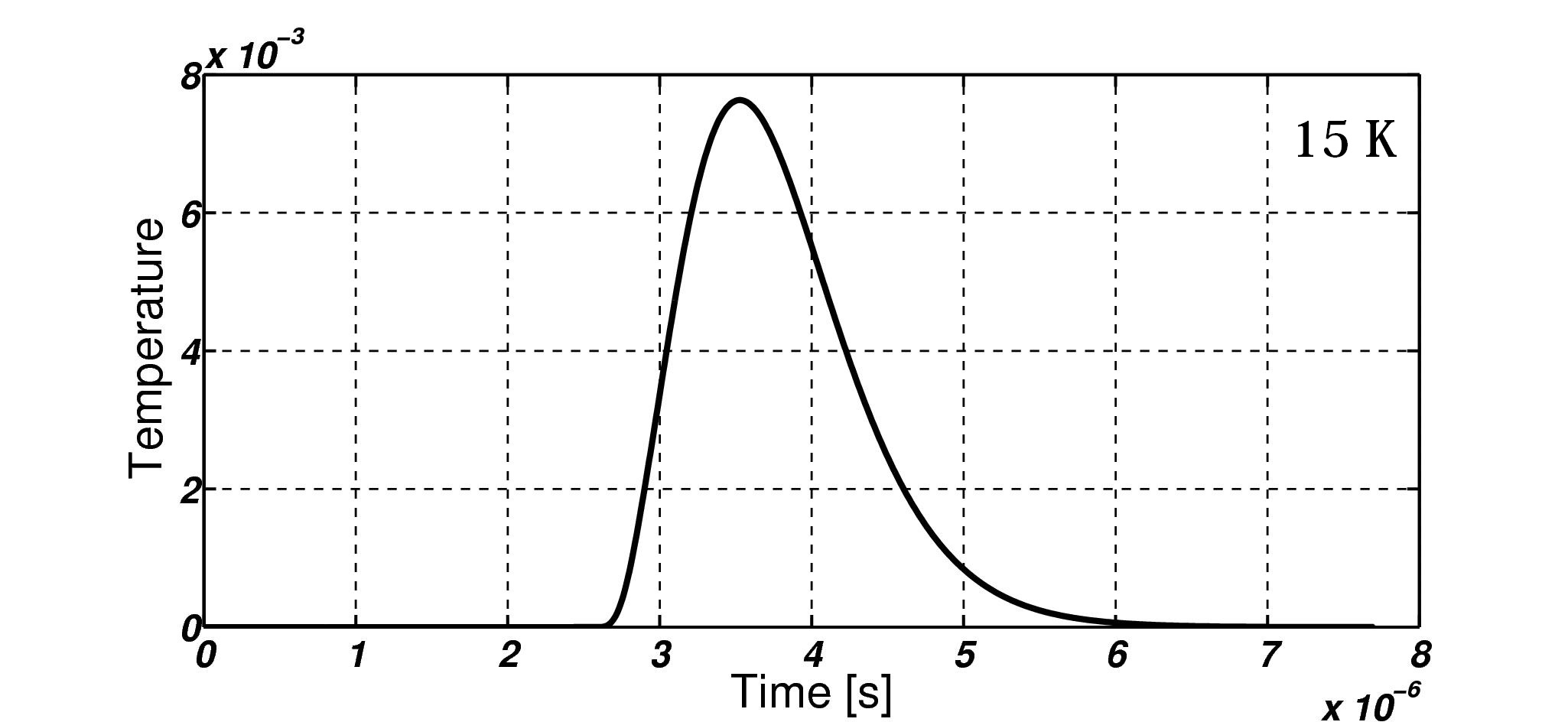}
         }% 
       \\  \subfigure[Related curve: @17.3K, sample \#7204205W]{%
           \label{fig:naf_sim_7}
            \includegraphics[width=7.5cm,height=4cm]{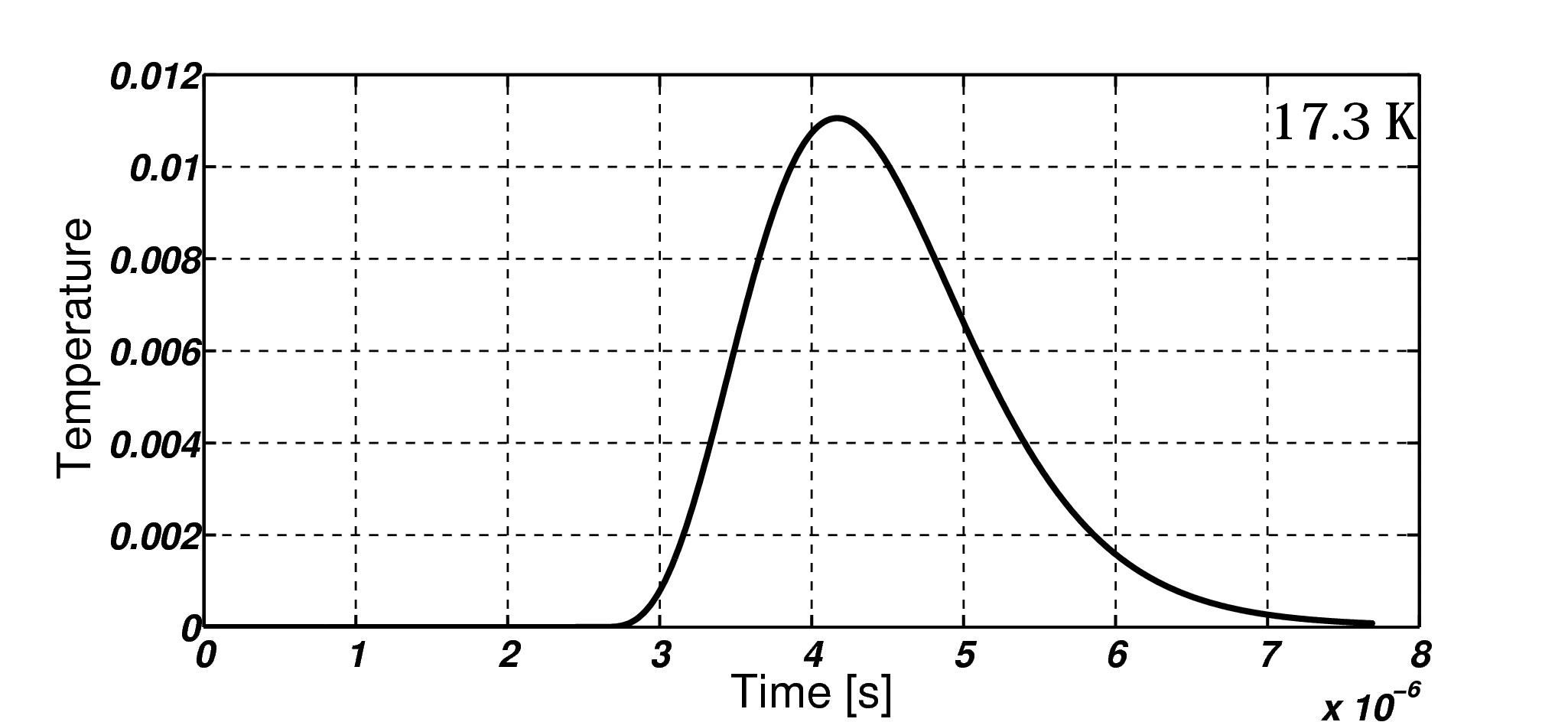}
         }% 
    \end{center}
    \caption{%
        Results of the simulations related to both crystals
     }%
   \label{fig:naf_simsall}
\end{figure}

\begin{figure}[H]
     \begin{center}
        \subfigure[Summarized results of simulation and measurement, the red curves corresponds to our calculations.]{%
            \label{fig:naf_fin}
            \includegraphics[width=8.5cm,height=7.5cm]{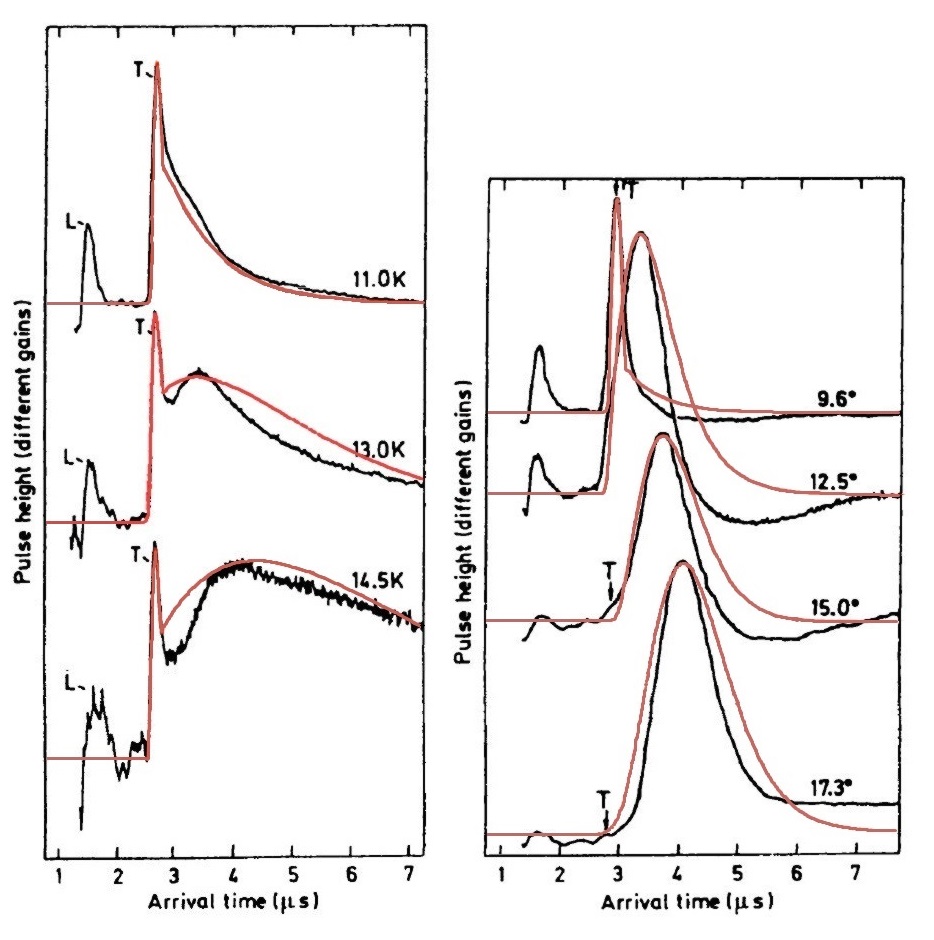}
        }%
    \\    \subfigure[Temperature dependence of the relaxation times related to sample \#607167J]{%
            \label{fig:naf1_reltim}
            \includegraphics[width=7cm,height=5cm]{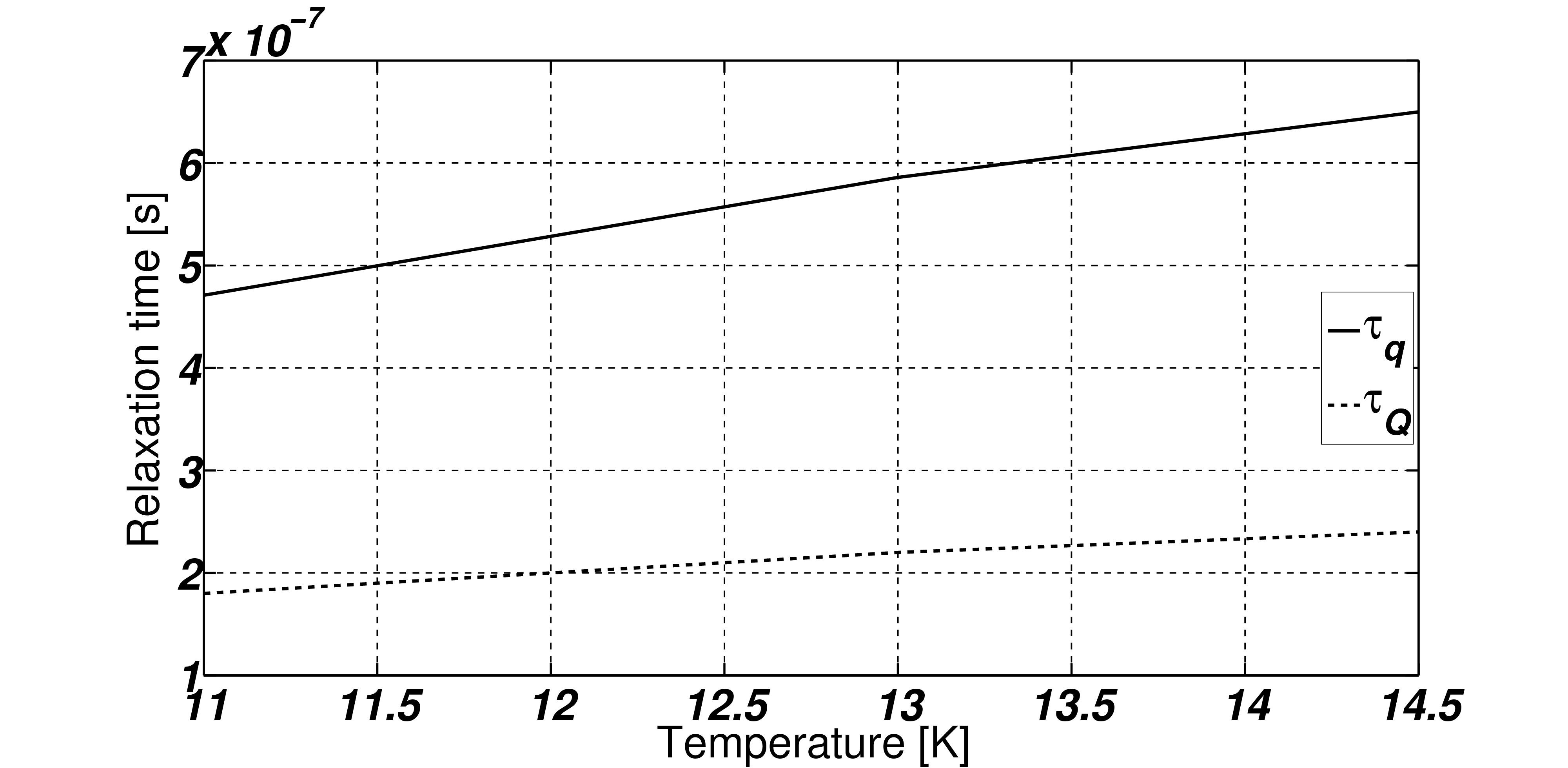}
        }%
        \subfigure[Related to sample \#607167J]{%
           \label{fig:naf1_HTC}
           \includegraphics[width=7cm,height=5cm]{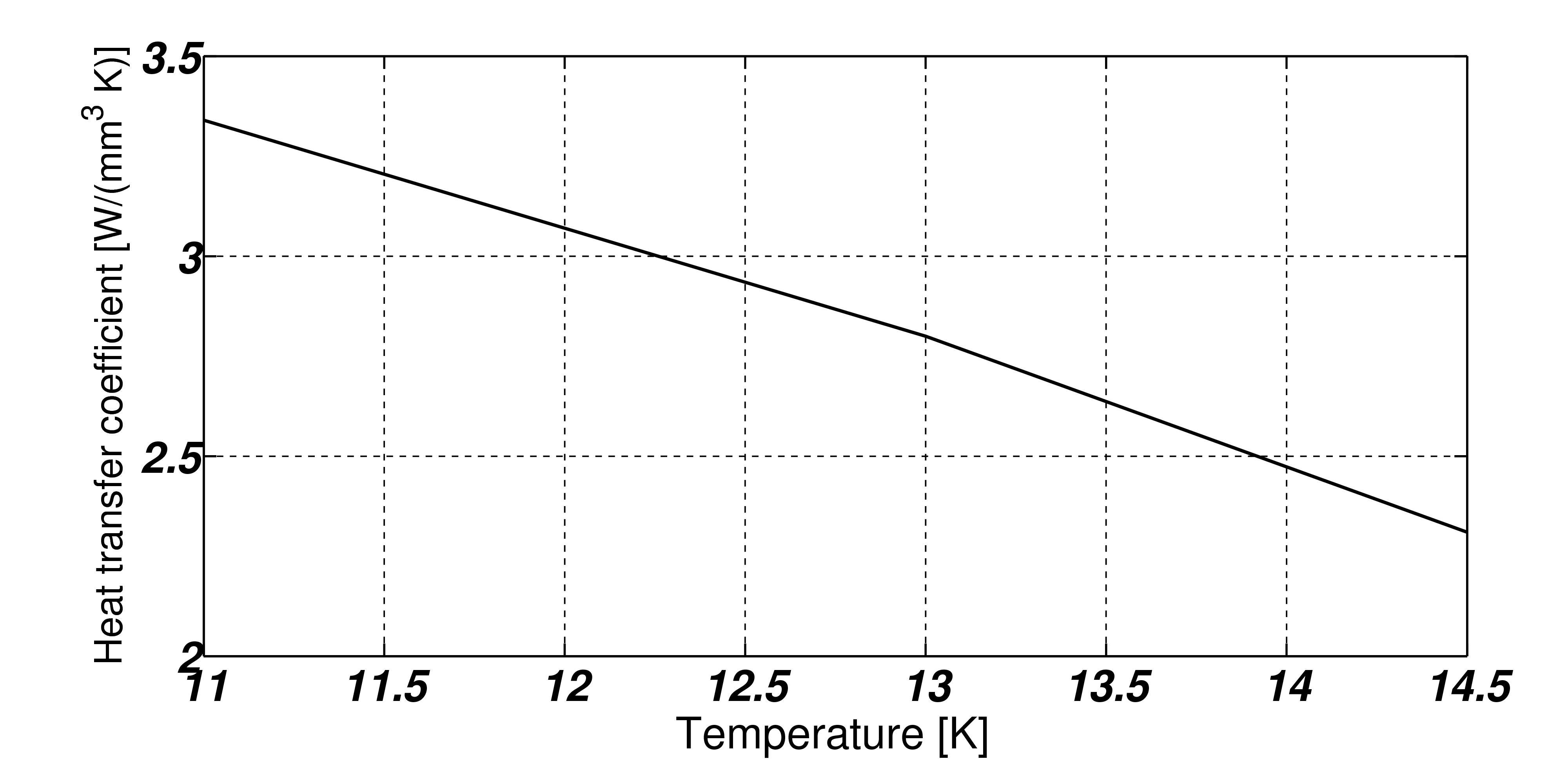}
        } %
        \subfigure[Temperature dependence of the relaxation times related to sample \#7204205W]{%
            \label{fig:naf2_reltim}
            \includegraphics[width=7cm,height=5cm]{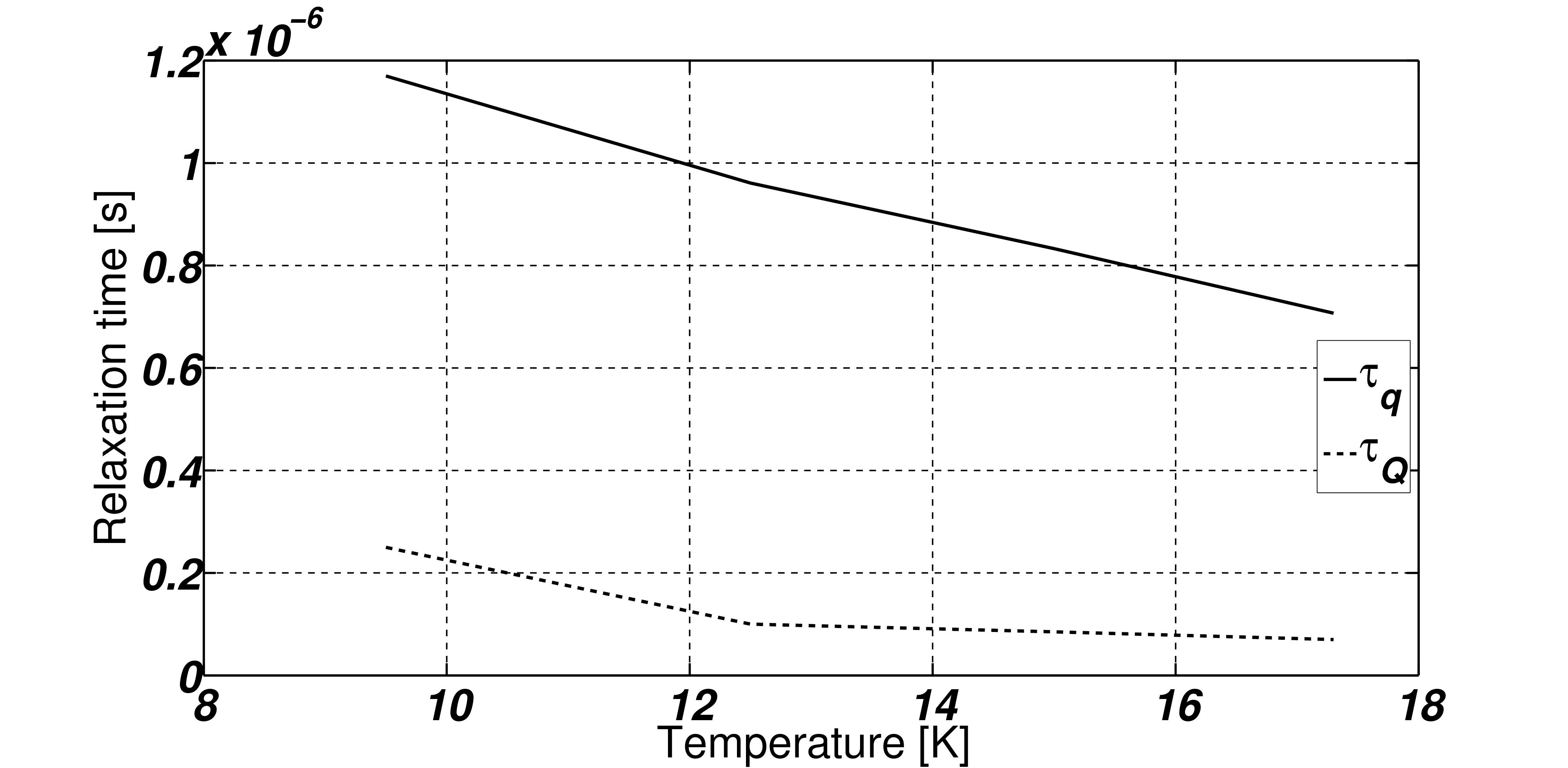}
        }%
         \subfigure[Related to sample \#7204205W]{%
            \label{fig:naf2_HTC}
            \includegraphics[width=7cm,height=5cm]{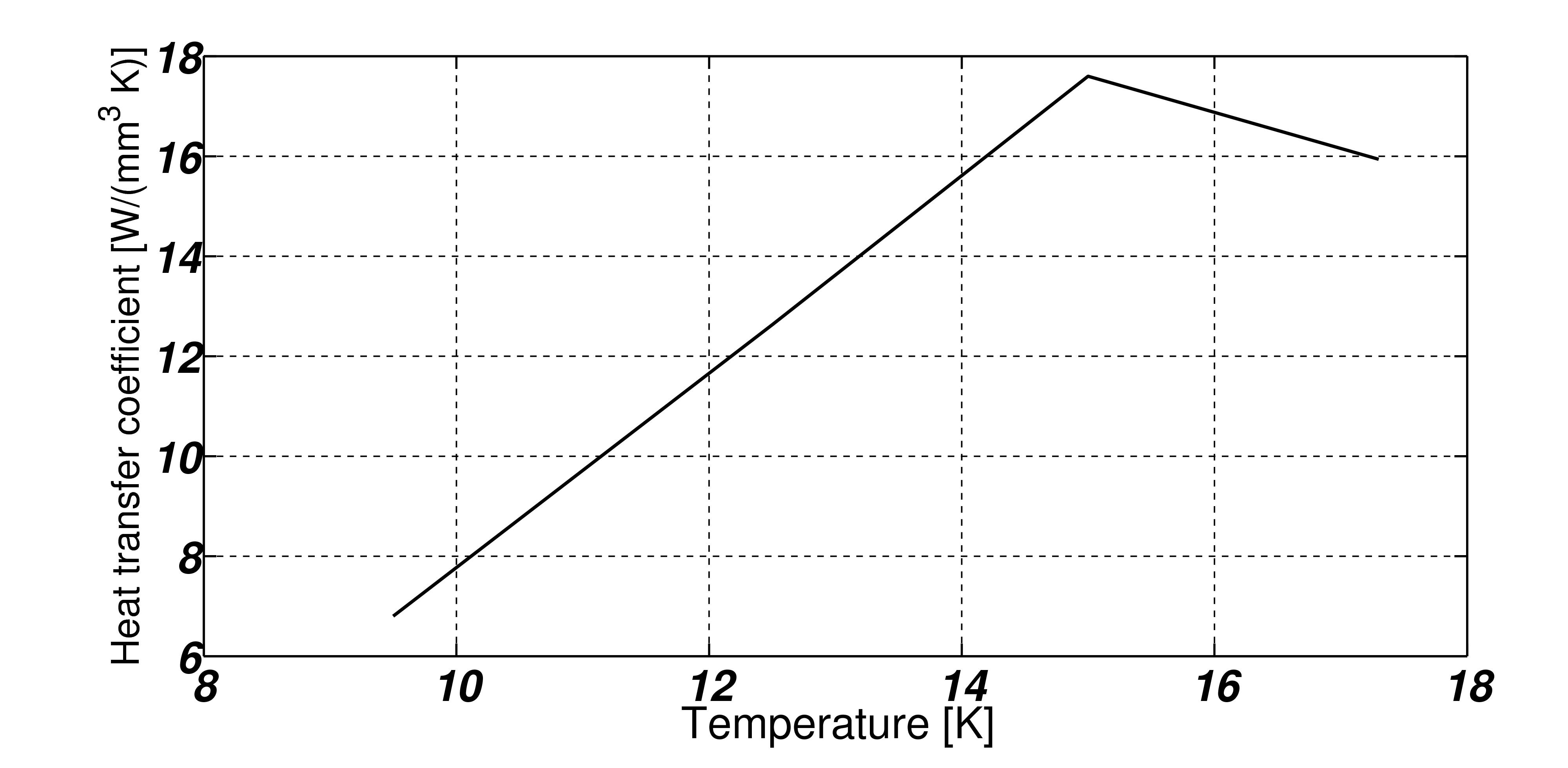}
        }
    \end{center}
    \caption{%
        The temperature dependence of the fitted parameters for different crystals
     }%
   \label{fig:naf_tempdeppar}
\end{figure}

One can see the results of the simulations (Figs. \ref{fig:naf_sim_1} - \ref{fig:naf_sim_7}) for each particular curves of Fig. \ref{fig:bal1}. There are two different side-effects of these results. The first one corresponds to the sample \#607167J. There is a broadening effect between the second sound and ballistic signal. The second one effects the other sample. Here, seemingly, the temperature goes below the initial temperature and the origin of this anomalous effect remains unknown because of the lack of informations regarding the precise experimental conditions.
Fig. \ref{fig:naf_fin} sums up all of the calculations and compares to the original curves.
The fitted parameters can be found in Tables \ref{naf:fitpar1} and \ref{naf:fitpar2}. Fig. \ref{fig:naf_tempdeppar} shows their temperature dependence. The goodness of this fitting can be determined by direct comparison with the results of Dreyer and Struchtrup \cite{DreStr93a} and Y. Ma  \cite{Ma13a1, Ma13a}.

\begin{table}[H]
\centering
\begin{tabular}{c|c|c|c}
       &Relax. time I. ($\tau_q$) [$\mu s$]&Relax. time II. ($\tau_Q$) [$\mu s$]  & \multicolumn{1}{c}{\begin{tabular}[c]{@{}c@{}}Heat transfer \\ coeff. ($a$) $\left [\frac{W}{mm^3K} \right ]$\end{tabular}}  \\ \hline
@11 K & 0.471 & 0.18 & 3.34  \\ \hline
@13 K &0.586 & 0.22 & 2.8\\ \hline
@14.5K &0.65& 0.24&2.31\\
    \end{tabular}\\
\caption{The fitted parameters for crystal \#607167J}
\label{naf:fitpar1}
\end{table}

\begin{table}[H]
\centering
\begin{tabular}{c|c|c|c}
       &Relax. time I. ($\tau_q$) [$\mu s$]&Relax. time II. ($\tau_Q$) [$\mu s$]  &\multicolumn{1}{c}{\begin{tabular}[c]{@{}c@{}}Heat transfer \\ coeff. ($a$) $\left [\frac{W}{mm^3K} \right ]$\end{tabular}}  \\ \hline
@9.6 K & 1.17& 0.25 & 6.8 \\ \hline
@12.5 K &0.961& 0.1 & 12.63 \\ \hline
@15 K &0.833 & 0.085 & 17.6\\ \hline
@17.3 K &0.707& 0.07 & 15.94 \\
    \end{tabular}\\
\caption{The fitted parameters for crystal \#7204205W}
\label{naf:fitpar2}
\end{table}

%-----------------------------------------------------------------
%\begin{comment}
%\begin{landscape}

%\end{landscape}

\subsection{Comparison with the Rational Extended Thermodynamical (RET) model}
\begin{landscape}

\begin{figure}[t!]
     \begin{center}
        \subfigure[Related crystal: \#607167J]{%
            \label{fig:naf_mulrug1}
            \includegraphics[width=9cm,height=12cm]{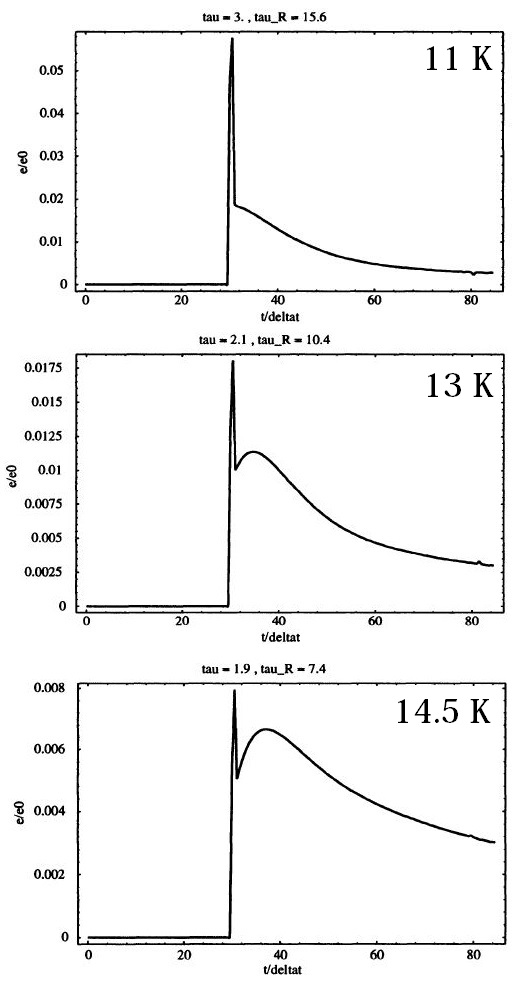}
        }%
        \subfigure[Related crystal: \#7204205W]{%
           \label{fig:naf_mulrug2}
           \includegraphics[width=16cm,height=12cm]{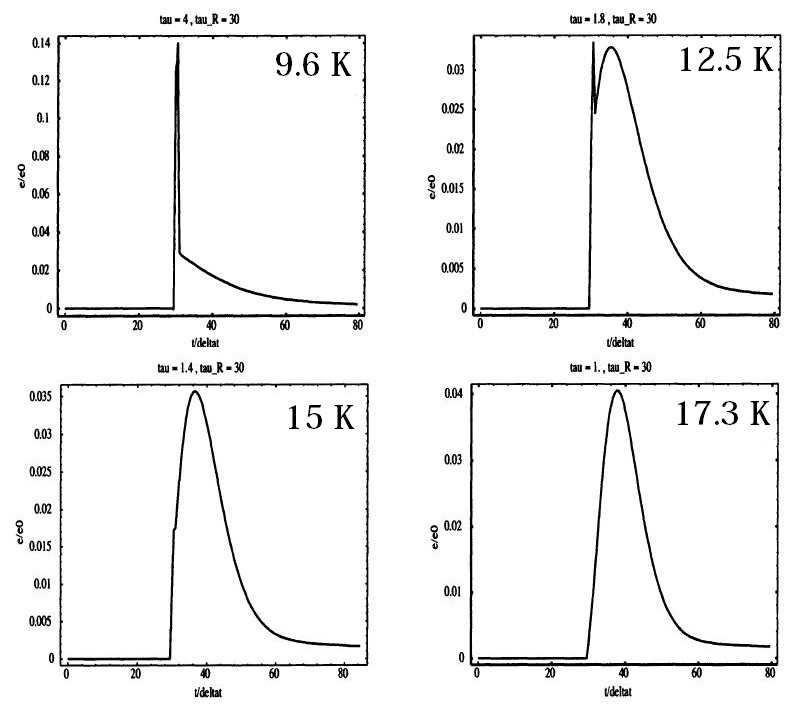}
        }

    \end{center}
    \caption{%
        Summarized result of the simulations from phonon hydrodynamical model with three momentum equations \cite{MulRug98}
     }%
   \label{fig:naf_all_mulrug}
\end{figure}

\end{landscape}

In the calculations of Dreyer and Struchtrup \cite{DreStr93a} the ballistic propagation speed is different from the measured value. However, in this RET based phonon hydrodynamical model only two relaxation time parameters are to be fitted. The result of the original calculations can be seen on Fig. \ref{fig:naf_all_mulrug}.

It is worth to compare the temperature dependency of relaxation times. The fitted relaxation times from RET are summarized in the Tables \ref{naf:fitpar_mulrug1} and \ref{naf:fitpar_mulrug2} below based on \cite{MulRug98}. The correspondence between the RET and BC models is:
\begin{equation*}
\tau_q = \tau_R, \quad
\tau_Q = \tau. \quad
\end{equation*}

\begin{table}[H]
\centering
\begin{tabular}{c|c|c}
       &Relax. time I. ($\tau_q$) [$\mu s$]&Relax. time II. ($\tau_Q$) [$\mu s$]     \\ \hline
@11 K & 1.56 & 0.3  \\ \hline
@13K & 1.04 & 0.21 \\ \hline
@14.5K &0.74& 0.19 \\
    \end{tabular}\\
\caption{The fitted parameters for crystal \#607167J}
\label{naf:fitpar_mulrug1}
\end{table}

\begin{table}[H]
\centering
\begin{tabular}{c|c|c}
       &Relax. time I. ($\tau_q$) [$\mu s$]&Relax. time II. ($\tau_Q$) [$\mu s$]     \\ \hline
@9.6 K & 3 & 0.4  \\ \hline
@12.5 K & 3 & 0.18 \\ \hline
@15 K &3& 0.14 \\ \hline
@17.3 K &3& 0.1 \\
    \end{tabular}\\
\caption{The fitted parameters for crystal \#7204205W}
\label{naf:fitpar_mulrug2}
\end{table}

It is remarkable to note that the $\tau_q$ time is constant in case of crystal \#7204205W. Moreover, the tendency of temperature dependence is opposite for crystal \#607167J comparing to the BC model. Beside the inappropriate value of thermal conductivity, the relative amplitudes seem to be also inaccurate. Moreover, a clear ballistic signal at $12.5$ K is predicted by the theory but it does not exists in the experiment.

\subsection{Comparison with the hybrid phonon gas model}

The difference between the prediction and the measured results is considerably higher in case of Y. Ma's model \cite{Ma13a, Ma13a1}. Tables \ref{naf:fitpar_yma1} and \ref{naf:fitpar_yma2} summarize the fitted relaxation time parameters. The values of $\tau_R (=\tau_q)$ and $\tau_N$ are given in \cite{Ma13a1} and $\tau (=\tau_Q)$ must be calculated according to the kinetic theory:
\begin{equation}
\f{1}{\tau} = \f{1}{\tau_R} + \f{1}{\tau_N}.
\end{equation}

\begin{figure}[H]
\centering
\includegraphics[width=11cm,height=5.5cm]{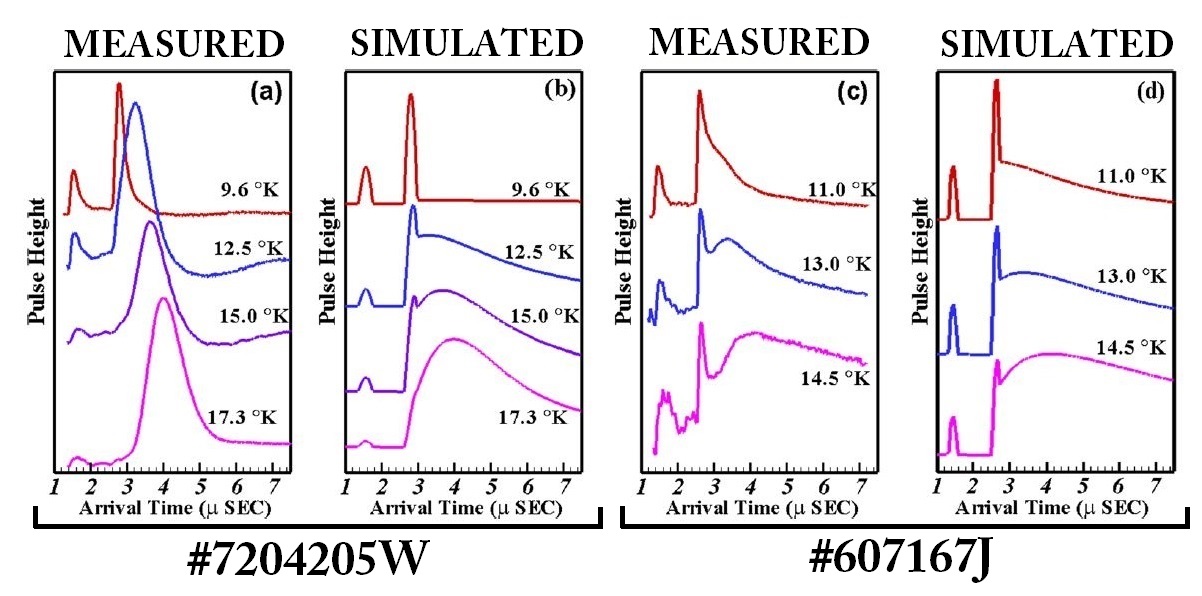}
\caption{Summarized result of the simulations from hybrid phonon gas model \cite{Ma13a1}}
\label{fig:naf_yma1}
\end{figure}

Fig. \ref{fig:naf_yma1} compares the theoretical results to the experiments.
In case of crystal \#7204205W $@9.6$ K, there is no clear sign of the second sound thus Ma adjusted $\tau_N$ as infinite, i.e. $\tau_R = \tau$. Moreover, a clear analogy can be observed with the results of Dreyer and Struchtrup \cite{DreStr93a}. The temperature dependency seems to be the same in these cases.
It is remarkable that in the papers of Y. Ma \cite{Ma13a, Ma13a1} the boundary conditions are missing.

\begin{table}[H]
\centering
\begin{tabular}{c|c|c}
       &Relax. time I. ($\tau_q$) [$\mu s$]&Relax. time II. ($\tau_Q$) [$\mu s$]     \\ \hline
@11 K & 1.056 & 0.281  \\ \hline
@13K & 0.937 & 0.248 \\ \hline
@14.5K &0.723& 0.208 \\
    \end{tabular}\\
\caption{The fitted parameters for crystal \#607167J}
\label{naf:fitpar_yma1}
\end{table}

\begin{table}[H]
\centering
\begin{tabular}{c|c|c}
       &Relax. time I. ($\tau_q$) [$\mu s$]&Relax. time II. ($\tau_Q$) [$\mu s$]     \\ \hline
@9.6 K & 1.56 & 1.56  \\ \hline
@12.5 K & 1.56 & 0.294 \\ \hline
@15 K &1.56& 0.245 \\ \hline
@17.3 K &1.56& 0.17 \\
    \end{tabular}\\
\caption{The fitted parameters for crystal \#7204205W}
\label{naf:fitpar_yma2}
\end{table}

\section{Summary}

The NaF experiments of second sound and ballistic phonon propagation performed by Jackson et al. are quantitatively analyzed in this paper. Performance of the ballistic-conductive equations of non-equilibrium thermodynamics with internal variables  is compared to the performance of the 3 field equations of Rational Extended Thermodynamics (RET) by Dreyer and Struchtrup \cite{DreStr93a} and to the ones based on hybrid phonon gas model by Y. Ma \cite{Ma13a, Ma13a1}. The effectiveness of the ballistic-conductive model is outstanding and competitive with both models. 

The differences are important not only in the performance, in concepts, but also in qualitative properties. E.g. it is shown that the relaxation time $\tau_q$ have a temperature dependency in non-equilibrium thermodynamics but in RET it has not. It is also important that neither RET, nor the complex viscosity approach consider cooling at the rear side, in spite of its inevitable appearance in the experiments.

The different conceptual background and the difference in validity of the theories are remarkable, too. Complex viscosity theory introduces a hydrodynamic equation for energy propagation, and complex viscosity simulates an additional damping effect over the viscosity. This suggestion is analogous to thermo-mass theory of Guo \cite{Guo06,GuoHou10, Guo2010general} and lacks a fundamental background. The version of phonon hydrodynamics by Rational Extended Thermodynamics has a definite microscopic background and fixes the particular parameters. In principle it is a theory of rarefied gases, the particular closure influences the validity range. The recent poor performance in modelling could be improved by considering the new dense gas extension of Ruggeri, Arima and Sugiyama \cite{RugSug15b, Arietal12, Arietal13, Arietal15, PenRugg16} or a better understanding of the internal mechanisms beyond the simple two channel relaxation of the Callaway collision integral. However, the necessity of large number of equations with increasing tensorial orders looks like an important practical and theoretical drawback. The non-equilibrium thermodynamic theory is compatible with the equations of the kinetic theory and has a flexibility of modelling, therefore its performance was the best. The validity of the theory is not restricted by particular microscopic pictures, it is based on the second law only \cite{KovVan15}. Therefore the theory does not exclude ballistic propagation of heat at room temperature, due to  material heterogeneities as an origin of a particular heat conduction mechanism \cite{PMMar17conf, Mari14, BerVan17b}. 

The available experimental data are old and some conditions indicate possible crucial problems. E.g. in Fig. \ref{fig:McN1} the thermal excitation is point-like compared to the size of the sample. Therefore a one dimensional propagation is not ensured and two dimensional effects are expected. However, due to the lack of related information and parameters a quantitative theoretical analysis is meaningless. In order to improve our understanding of wave like propagation modes of heat and to develop the corresponding exciting technology of dynamically regulated heat conduction new experimental data are necessary. 

\section{Acknowledgements}
The work was supported by the grant OTKA K116197.



 % \bibliographystyle{elsarticle-num} 
% \bibliography{references}


%\begin{thebibliography}{00}


%\bibitem{}

%\end{thebibliography}
\end{document}